\documentclass[10pt, twocolumn, letterpaper]{article}

\usepackage{cvpr}              

\makeatletter
\providecommand{\@LN}[2]{}
\makeatother

\definecolor{cvprblue}{rgb}{0.21,0.49,0.74}
\usepackage[pagebackref, breaklinks, colorlinks, allcolors=cvprblue]{hyperref}
\usepackage{hhline}
\usepackage{colortbl}
\usepackage{nicefrac}
\usepackage{multirow}
\usepackage{makecell}
\usepackage{algorithmic, algorithm}
\definecolor{shadegray}{RGB}{211,211,211}
\definecolor{lightgray}{RGB}{230,230,230}
\graphicspath{{figure/}}

\def\paperID{16447} 
\def\confName{CVPR}
\def\confYear{2025}

\title{Image Quality Assessment: Enhancing Perceptual Exploration and Interpretation with Collaborative Feature Refinement and Hausdorff distance}

\author{Xuekai Wei, Junyu Zhang, Qinlin Hu, Mingliang Zhou\\Yong Feng, Weizhi Xian, Huayan Pu\\
Chongqing University\\
Chongqing 400044, China\\
{\tt\small xuekaiwei2-c@my.cityu.edu.hk, 202414021003@stu.cqu.edu.cn}\\
{\tt\small huqinlin@stu.cqu.edu.cn, mingliangzhou@cqu.edu.cn}\\
{\tt\small fengyong@cqu.edu.cn, wasxxwz@163.com}\\
{\tt\small phygood\_2001@shu.edu.cn}\\
\and
Sam Kwong\\
Lingnan University\\
Hong Kong 999077\\
{\tt\small samkwong@ln.edu.hk}
}

\begin{document}
\maketitle
\begin{abstract}
Current full-reference image quality assessment (FR-IQA) methods often fuse features from reference and distorted images, overlooking that color and luminance distortions occur mainly at low frequencies, whereas edge and texture distortions occur at high frequencies. This work introduces a pioneering training-free FR-IQA method that accurately predicts image quality in alignment with the human visual system (HVS) by leveraging a novel perceptual degradation modelling approach to address this limitation. First, a collaborative feature refinement module employs a carefully designed wavelet transform to extract perceptually relevant features, capturing multiscale perceptual information and mimicking how the HVS analyses visual information at various scales and orientations in the spatial and frequency domains. Second, a Hausdorff distance-based distribution similarity measurement module robustly assesses the discrepancy between the feature distributions of the reference and distorted images, effectively handling outliers and variations while mimicking the ability of HVS to perceive and tolerate certain levels of distortion. The proposed method accurately captures perceptual quality differences without requiring training data or subjective quality scores. Extensive experiments on multiple benchmark datasets demonstrate superior performance compared with existing state-of-the-art approaches, highlighting its ability to correlate strongly with the HVS.\footnote{The code is available at \url{https://anonymous.4open.science/r/CVPR2025-F339}.}
\end{abstract}

\section{Introduction}
\label{sec:intro}
The development of reliable and consistent image quality assessment (IQA) methods is essential for training and optimizing computer vision models to improve their performance in practical applications~\cite{8796396}. IQA methods can be divided into subjective and objective IQA methods depending on the participation of human observers~\cite{9932271}. In subjective quality assessment, observers evaluate image samples on the basis of personal opinions, which can lead to variations in quality scores for the same image, requiring highly skilled participants~\cite{10180056}. They are time-consuming and costly to organize, and objective evaluation has become a common alternative. The objective IQA methods simulate human visual system (HVS) behavior by learning image features, predicting the perceived quality of images, and matching the human subjective mean opinion score (MOS) or differential quality score (DMOS)~\cite{10409615}, which can be categorized as full-reference (FR)-IQA models~\cite{Ding2021ComparisonOF}, reduced-reference (RR)-IQA models~\cite{10014662}, and no-reference (NR)-IQA models~\cite{NEURIPS2024_ni,10539337}, depending on their reliance on information from the original reference image. FR-IQA methods require access to the original reference image to assess the quality of a processed image, aligning with the HVS ability to more accurately perceive the degree of quality degradation in processed images compared with the original method~\cite{10.1145/3474085.3475419}. FR-IQA methods often provide more reliable quality assessments, which can guide various neural visual models as perceptual losses, such as image denoising~\cite{Chen_2023_CVPR,9917526}, superresolution~\cite{Tsao_2024_CVPR,Chen_2024_CVPR}, and compression~\cite{Jia_2024_CVPR,Xing_2024_CVPR}.

FR-IQA methods typically evaluate the quality of a distorted image by measuring the distance between the reference and distorted images~\cite{10.1145/3503161.3548193}. Traditional FR-IQA approaches are based on modelling key features of the HVS, such as sensitivity to brightness, chroma, and frequency content of visual stimuli~\cite{10.1145/3447393}. While these methods aim to simulate human perception, they are subject to limitations due to their assumptions about the HVS and often perform poorly in practical image quality assessment tasks~\cite{10478301}. Deep feature extraction and comparison have heralded a powerful paradigm shift in the FR-IQA methods introduced by the deep learning-based image quality assessment framework~\cite{8099696}. Neural-enhanced FR-IQA models employ neural networks, such as the VGG network~\cite{Shukla_2024_WACV}, the squeeze network~\cite{10.1145/3343031.3350990}, and the Alex network~\cite{10.1007/978-3-030-58621-837}, to extract perceptual features, yielding remarkable results across various IQA datasets~\cite{Zhang_2023_CVPR}. Furthermore, recent advanced models successfully generate perceptual quality scores by leveraging the statistical distribution distance to compare the extracted features, demonstrating unparalleled performance on conventional and texture image datasets~\cite{GIANNITRAPANI2025117212}. The rapid advancements in deep learning-based approaches to FR-IQA have offered many opportunities for exploring novel techniques and pushing the boundaries of visual evaluation and optimization across a broad spectrum of applications.

However, existing FR-IQA methods often rely on pairwise comparisons of deep features from reference and distorted images, overlooking the crucial distinction that distortions in color and luminance primarily appear at low frequencies. In contrast, distortions in edges and textures are present mainly at high frequencies~\cite{WANG2023109296}. Furthermore, these methods are inflexible because they rely on human-labelled opinion scores and their design for evaluating fixed image characteristics with limitations of point-by-point distances in adequately representing the distances between shifted distortions~\cite{LAN2025121557}. While learning-based quality metrics have demonstrated impressive performance in various domains, they highly rely on the quality of the training data. The development of a genuinely generic and training-free FR-IQA model remains an open challenge because of the complexity and diversity of image distortions and the subjective nature of human perception.

To address these challenges, we introduce a novel and comprehensive, objective quality assessment model to provide a more robust and accurate evaluation of image quality across a diverse range of synthesized images and distortions. Our approach harnesses the power of the discrete wavelet transform (DWT)~\cite{10409615} to effectively capture and represent the intricate features of image distortions. Inspired by the hierarchical property of the HVS in perceiving the visual world, we incorporate a multiscale DWT scheme into our model. This multiscale approach enables us to precisely localize and quantify local distortions in natural images by leveraging a combination of techniques, including visual saliency, region of interest extraction, morphological opening operations, thresholding, and filtering. The proposed model aligns closely with human perception by focusing on perceptually essential features, providing a more reliable and accurate image quality assessment. This paper proposes a training-free approach that enhances generalizability and practicality.

\begin{itemize}
\item We propose a novel training-free full-reference image quality assessment method that leverages perceptual-related domain transformation and distribution similarity measurement to accurately predict image quality aligned with human visual perception. The proposed method effectively captures perceptual quality differences between reference and distorted images through the integration of wavelet transform-based feature extraction and Hausdorff distance-based distribution comparison.
\item The proposed wavelet domain assessment capability is a key strength, as it effectively captures both overall image quality and localized distortions. The wavelet transform-based domain transformation provides a multiscale representation of the image, enabling the extraction of perceptually relevant features at different levels of detail.
\item Simultaneously, the Hausdorff distance-based distribution similarity measurement is sensitive to local distortions and outliers, ensuring that the method can accurately identify and quantify local quality degradation. The Hausdorff distance is sensitive to outliers, so it can effectively capture the presence of local distortions or artifacts in the distorted image, making it suitable for detecting and quantifying local quality degradations that may significantly affect perceived image quality.
\end{itemize}
\section{Related Work}
Commonly established FR-IQA algorithms first standardize the resolution of the original reference image and distorted image and then calculate their similarity scores through feature extraction and distance judgment~\cite{10.1145/3447393}. Traditional methods such as the mean square error (MSE)~\cite{6562752} and peak signal-to-noise ratio (PSNR)~\cite{8980171} are commonly used but have limited effectiveness, ignoring the HVS perspective. Therefore, structural similarity (SSIM)~\cite{1284395}, its extended multiscale version (MS-SSIM)~\cite{1292216}, and the feature similarity index for image quality assessment (FSIM)~\cite{5705575} were introduced to better capture luminance, contrast, and structural features. Additionally, convolutional neural networks (CNNs) excel at extracting multiscale image features and have become central to learning-based IQA algorithms~\cite{8099696}. In addition, algorithms such as LPIPS~\cite{8578166} and DISTS~\cite{9298952} use CNNs to extract multilevel features for distance calculation. The transformer is also used for feature extraction in attention-based FR-IQA networks~\cite{9523022}, whereas ensemble models such as IQMA~\cite{9523177} and EGB~\cite{9522809} enhance evaluation accuracy by averaging prediction scores, demonstrating the effectiveness of deep learning in image quality assessment. DeepWSD~\cite{10577432} introduces the statistical distribution distance to capture the pixel correlations among deep features, which correlate exceptionally well with the HVS on several standard IQA datasets~\cite{8743252}.

In computer vision, similarity judgment is crucial in various tasks, including image quality assessment, point cloud completion, and image retrieval. These tasks focus on establishing distance metrics and image descriptor-based loss functions~\cite{10577432}. Kullback‒Leibler divergence (KLD)~\cite{10.1117/12.597306} and Jensen‒Shannon divergence (JSD)~\cite{7559777} are widely used to compare image features across various visual tasks, including RR-IQA, NR-IQA, and image retrieval. Shape-level distances, such as the chamfer distance (CD)~\cite{NEURIPS2021_f3bd5ad5}, Earth Mover's distance (EMD)~\cite{Zhang_Zhang_Chan_Zhang_U_2024}, and the improved density-aware chamfer distance (DCD)~\cite{wu2021densityaware} or hyperbolic chamfer distance~\cite{Lin_2023_ICCV}, are widely used metrics in point cloud-related tasks. Image descriptors quantify the fundamental characteristics of images, such as texture, color, complexity, and quality. Deep features extracted by pretrained networks have paved the way for developing image descriptors that regress these features to scores~\cite{10716315}. Two notable similarity judgments are those of Hojjat \textit{et al.}~\cite{10533377} and Liao \textit{et al.}~\cite{10577432}, which employ the projected Wasserstein distance (WSD) measure to compare VGG features for image enhancement and for FR-IQA, respectively.

\section{Preliminaries}
The primary objective of FR-IQA is to develop a computational model that accurately predicts the perceived quality of a distorted image by comparing it with its corresponding pristine reference image. Let the reference image be denoted as $I_r$ and the distorted image be denoted as $I_d$~\cite{10577432}. The goal is to estimate a quality score $Q$ that quantifies the perceptual similarity between $I_r$ and $I_d$, where $Q$ is a continuous MOS-related value. A higher score represents perfect quality, indicating that the distorted image is perceptually indistinguishable from the reference image, whereas a lower score suggests severe distortion, meaning that the distorted image is significantly different from the reference image and exhibits highly noticeable and unpleasant artifacts. The FR-IQA problem can be mathematically formulated as $f$, which takes the reference image $I_r$ and the distorted image $I_d$ as inputs and maps them to a quality score~\cite{10.1117/12.597306,7559777}:
\begin{equation}
Q = f(I_r, I_d) \in [Q_{min}, Q_{max}]
\end{equation}

The function $f$ represents the quality assessment model that encapsulates the underlying computational mechanisms and algorithms used to compare the reference and distorted images and estimate the quality score. $Q_{min}$ and $Q_{max}$ are the minimum and maximum values of the MOS-related quality scale, respectively. The specific range of $Q$ depends on the chosen MOS scale, which can vary across different subjective quality assessment datasets and experiments. $Q_{min}$ can be formulated as follows~\cite{9298952}.
\begin{equation}
Q_{min} = \min_{I_d \in \mathcal{D}} f(I_r, I_d)
\end{equation}
where $\mathcal{D}$ represents the set of all possible distorted images that can be generated from the reference image $I_r$. $Q_{max}$ represents the ideal scenario with no perceptual difference between the distorted and reference images, indicating perfect quality. The commonly used $f(I_r, I_d)$ function can be formulated as the similarity measure $sim(I_r, I_d)$ of the distribution between the reference and the distorted images~\cite{1284395}.
\begin{equation}
sim(I_r, I_d) = \dfrac{1}{dist(F(I_r), F(I_d))}
\end{equation}
where $F(*)$ represents the feature representations of the reference and distorted images in the deep feature space. The primary goal of image quality assessment is to minimize the discrepancy between the predicted quality score and the subjective MOS, which can be formulated as follows:
\begin{equation}
\min_{f} \mathbb{E}[L(sim(I_r, I_d), MOS)]
\end{equation}
where $L(\cdot)$ is a loss function that quantifies the difference between the predicted quality score $f(I_r, I_d)$ and the subjective quality score $MOS$. The expectation $\mathbb{E}$ is taken over the data distribution, considering a wide range of reference images, distorted images, and their corresponding subjective quality scores. The ultimate goal of FR-IQA is to develop a robust and reliable quality assessment model $f$ that can accurately predict the perceptual quality of distorted images across a wide range of distortion types, image content, and viewing conditions. The model should be able to generalize well to unseen images and distortions, providing quality scores that closely align with human subjective opinions.
\begin{figure*}[!t]
\centerline{\includegraphics[width=0.96\linewidth]{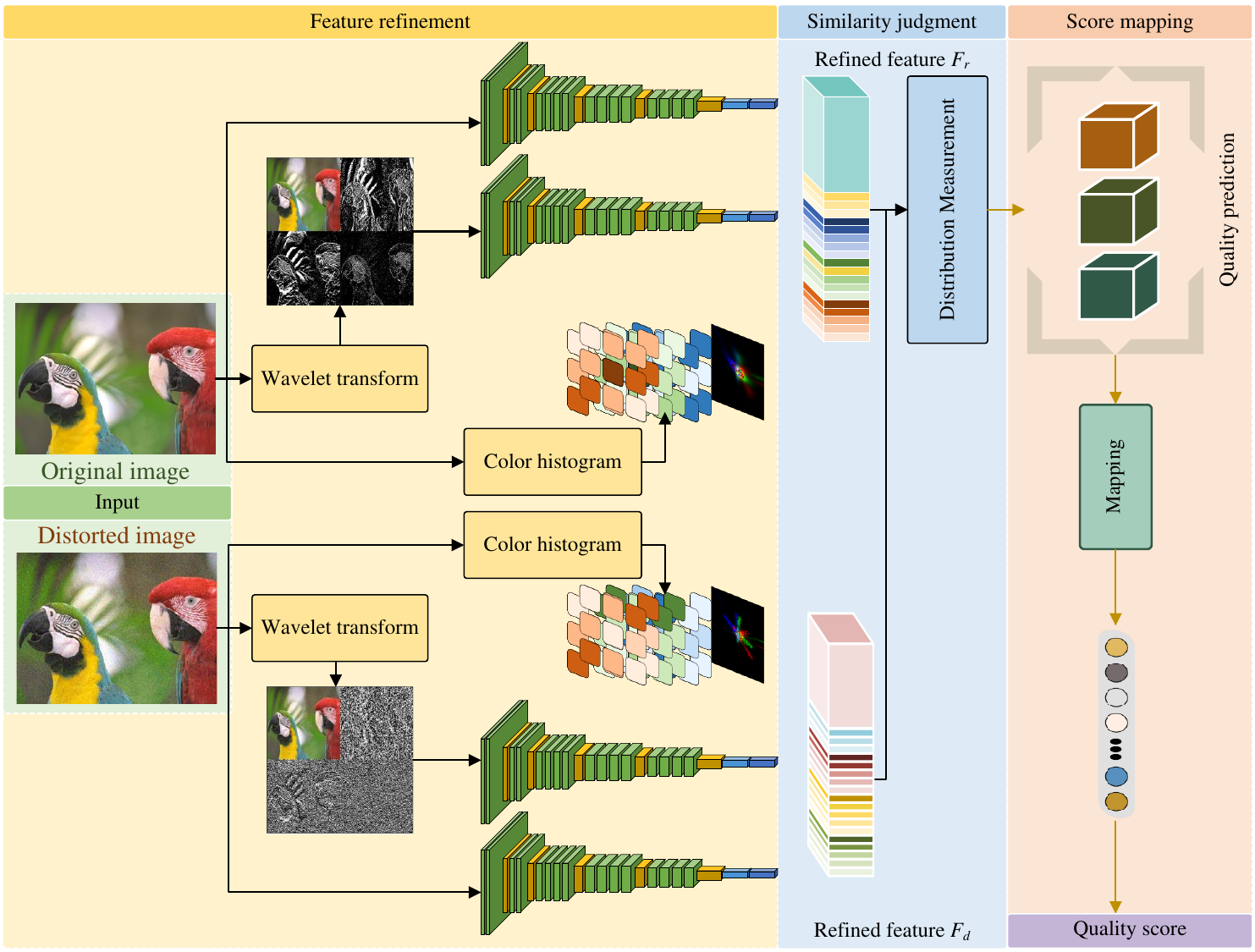}}
\caption{Framework of the proposed FR-IQA algorithm. }\label{fig01}
\end{figure*}
\section{Method}
\subsection{Framework}
The proposed FR-IQA method consists of a novel framework that effectively enhances perceptual exploration and interpretation by leveraging collaborative feature refinement and the Hausdorff distance. Fig. \ref{fig01} illustrates the overall framework of the proposed method for enhancing perceptual exploration and interpretation with collaborative feature refinement and the Hausdorff distance. The framework comprises two main components: a collaborative feature refinement module and a Hausdorff distance-based distribution similarity measurement module. The collaborative feature refinement module employs a carefully designed wavelet transform to extract perceptually relevant features from reference and distorted images. The wavelet transform decomposes the images into multiple frequency subbands, enabling the capture of perceptual information. This decomposition mimics how the HVS analyses visual information at various scales and orientations. By separating the low-frequency and high-frequency components, the module effectively captures the frequency-dependent nature of distortions, where color and luminance distortions predominantly manifest in low-frequency bands, whereas edge and texture distortions are primarily concentrated in high-frequency bands. The collaborative nature of this module allows for the refinement of features on the basis of their perceptual relevance and the characteristics of the distortions present in the image.

The refined features from the collaborative feature refinement module are then passed to the Hausdorff distance-based distribution similarity measurement module. This module robustly assesses the discrepancy between the feature distributions of the reference and distorted images. The Hausdorff distance is employed as a similarity metric, effectively handling outliers and variations in the feature distributions. By considering the entire distribution of features rather than relying on point estimates, the module captures the perceptual quality degradation in a manner consistent with the ability of HVS to perceive and tolerate certain levels of distortion while being sensitive to perceptually significant changes. The outputs from the collaborative feature refinement module and the Hausdorff distance-based distribution similarity measurement module are combined to generate the final perceptual quality score. This integration allows for a comprehensive assessment of image quality that considers both the frequency-dependent nature of distortions and the perceptual relevance of extracted features.

\subsection{Collaborative Feature Refinement Module}
In the collaborative feature refinement module, a wavelet transform is employed to extract perceptually relevant features from both the reference image $I_r$ and the distorted image $I_d$. The wavelet transform decomposes the images into multiple frequency subbands~\cite{10.1117/12.597306}. Let $W_r$ and $W_d$ denote the wavelet decompositions of $I_r$ and $I_d$, respectively:
\begin{equation}
W_r = \mathcal{W}(I_r),
W_d = \mathcal{W}(I_d)
\end{equation}
where $\mathcal{W}$ represents the wavelet transform operator. The wavelet decompositions $W_r$ and $W_d$ consist of low-frequency and high-frequency subbands, which effectively capture the frequency-dependent nature of distortions. The low-frequency subbands, denoted as $W_r^L$ and $W_d^L$, primarily contain information related to color and luminance distortions, whereas the high-frequency subbands, denoted as $W_r^H$ and $W_d^H$, capture edge and texture distortions. The collaborative feature refinement module then applies a refinement function $\mathcal{R}$ to the wavelet subbands, enhancing their perceptual relevance and adaptively weighting them on the basis of their importance in capturing distortions:
\begin{equation}
F_r = \mathcal{R}(W_r^L, W_r^H),
F_d = \mathcal{R}(W_d^L, W_d^H)
\end{equation}
where $F_r$ and $F_d$ are the refined feature representations of the reference and distorted images, respectively. The collaborative feature refinement module employs a wavelet transform-based approach to extract perceptually relevant features from the reference and distorted images. The wavelet transform is performed via four filters, denoted as $f_{LL}$, $f_{LH}$, $f_{HL}$, and $f_{HH}$, which represent the low-low, low-high, high-low, and high-high frequency components, respectively. The input image $Img$ is convolved with a set of filters ${f_{LL}, f_{LH}, f_{HL}, f_{HH}}$ to obtain four subband images~\cite{10409615}:
\begin{equation}
\begin{split}
&[S_{LL}, S_{LH}, S_{HL}, S_{HH}] = \\
&\operatorname{Conv}([f_{LL}, f_{LH}, f_{HL}, f_{HH}], Img)
\end{split}
\end{equation}

The subband images are then decomposed into approximation coefficients $C_A$ and detail coefficients $C_D$:
\begin{equation}
\begin{cases}
C_A(i, j) = \frac{S(i, 2j-1) + S(i, 2j)}{2} \\
C_D(i, j) = S(2i-1, j) - S(2i, j)
\end{cases}
\end{equation}
where $S$ represents the subband images $S_{LL}$, $S_{LH}$, $S_{HL}$, or $S_{HH}$. The approximation and detail coefficients are further decomposed into four components:
\begin{equation}
\begin{cases}
C_{AA}(i, j) = \frac{C_A(i, 2j-1) + C_A(i, 2j)}{2} \\
C_{AD}(i, j) = \frac{C_D(i, 2j-1) - C_D(i, 2j)}{2} \\
C_{DA}(i, j) = C_A(i, 2j-1) - C_A(i, 2j-1) \\
C_{DD}(i, j) = C_D(i, 2j-1) - C_D(i, 2j)
\end{cases}
\end{equation}
where $C_{AA}$, $C_{AD}$, $C_{DA}$, and $C_{DD}$ are the further decomposed components of the coefficients.
\subsection{Hausdorff Distance-based Distribution Similarity Measurement Module}
The refined feature representations $F_r$ and $F_d$ are then passed to the Hausdorff distance-based distribution similarity measurement module. This module assesses the discrepancy between the feature distributions of the reference and distorted images via the Hausdorff distance. Let $\mathcal{H}$ denote the Hausdorff distance operator. The similarity between the reference and distorted feature distributions can be computed as~\cite{9381694}:
\begin{equation}
sim_{Haus}(F_r, F_d) = MAP(\mathcal{H}(F_r, F_d))
\end{equation}

The Hausdorff distance effectively handles outliers and variations in the feature distributions, capturing the perceptual quality degradation consistently with the human visual system. The distribution similarity measurement module compares the distributions of the perceptually transformed features obtained from the reference and distorted images. The Hausdorff distance between two sets of features $F_r$ and $F_d$ is defined as:
\begin{equation}\begin{aligned}
&\mathcal{H}(F_r, F_d) = \\
&\max \left\{\sup_{f_r \in F_r} \inf_{f_d \in F_d} d(f_r, f_d), \sup_{f_d \in F_d} \inf_{f_r \in F_r} d(f_r, f_d)\right\}
\end{aligned}
\end{equation}

An upper bound for the Hausdorff distance in the feature space is given by:
\begin{equation}
\begin{aligned}
\mathcal{H}(F_r, F_d) &\leq \int_{F_r \times F_d} |f_r - f_d| d\gamma \\
&= \mathbb{E}_{(f_r, f_d) \sim \gamma} |f_r - f_d| \\
&= d_\gamma(F_r, F_d)
\end{aligned}
\end{equation}
where $d_\gamma(F_r, F_d)$ represents the expected distance between the feature distributions under the joint distribution $\gamma$.

To further enhance the perceptual relevance of the similarity measurement, the distance $ C(H_r, H_d) $ via the color histograms of the reference and distorted images is computed as the final similarity measure weight \cite{9577403}:
\begin{equation}
C(H_r, H_d) = \frac{1}{\sqrt{2}} \|H_r^{1/2} - H_d^{1/2}\|_2
\end{equation}
where $H_r$ and $H_d$ represent the histograms of the reference and distorted images, respectively, following the method proposed by Mahmoud \textit{et al.} \cite{9577403}.

The proposed method effectively integrates perceptual-related domain transformation and distribution similarity measurements to accurately predict image quality without relying on training data or subjective quality scores.


The final perceptual quality score $Q_p$ is estimated by combining the outputs from the collaborative feature refinement module and the Hausdorff distance-based distribution similarity measurement module:
\begin{equation}
Q_p = MAP_g(sim_{Haus}(F_r, F_d), C\left(\mathbf{H}_r, \mathbf{H}_d\right))
\end{equation}
where $MAP_g$ is a function that integrates the refined feature representations and the similarity score to produce the final quality score. The proposed framework enhances perceptual exploration and interpretation by leveraging collaborative feature refinement and the Hausdorff distance. The collaborative feature refinement module captures the frequency-dependent nature of distortions, whereas the Hausdorff distance-based distribution similarity measurement module robustly assesses the perceptual quality degradation. The integration of these two components results in a comprehensive and accurate assessment of image quality that aligns with human visual perception. The resulting quality prediction model can be directly applied to new test images to estimate their full-reference quality scores on the basis of the wavelet coefficient distributions without requiring a separate training phase.


\begin{figure*}[!t]
\centering
    \begin{subfigure}{0.3\linewidth}
        \includegraphics[width=\linewidth]{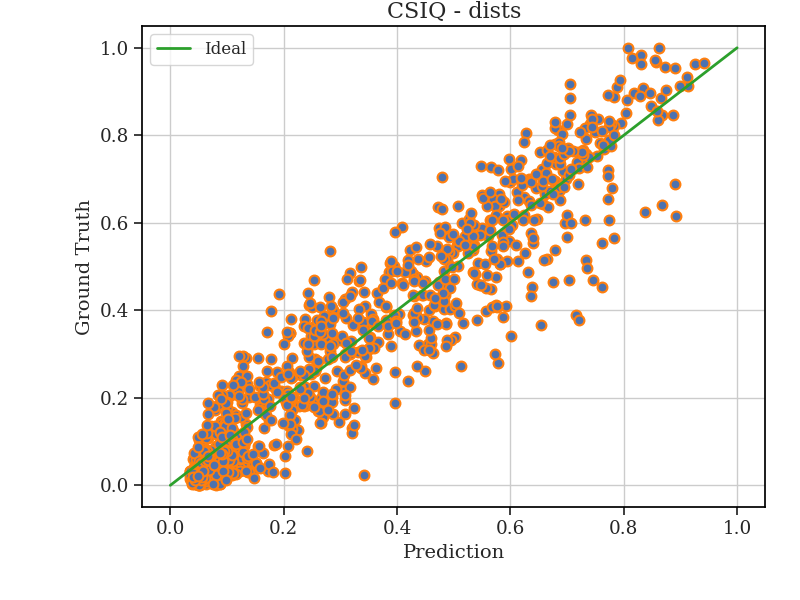}
    \end{subfigure} 
    \begin{subfigure}{0.3\linewidth}
        \includegraphics[width=\linewidth]{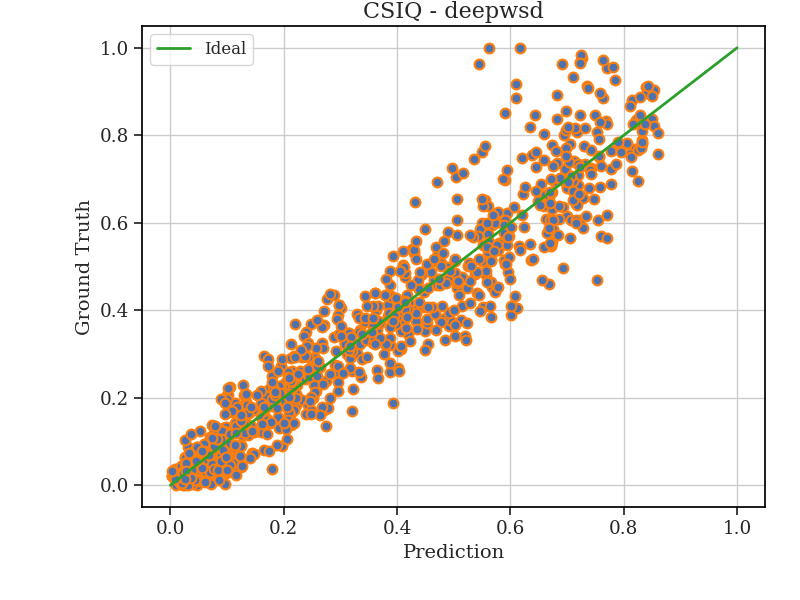}
    \end{subfigure} 
    \begin{subfigure}{0.3\linewidth}
        \includegraphics[width=\linewidth]{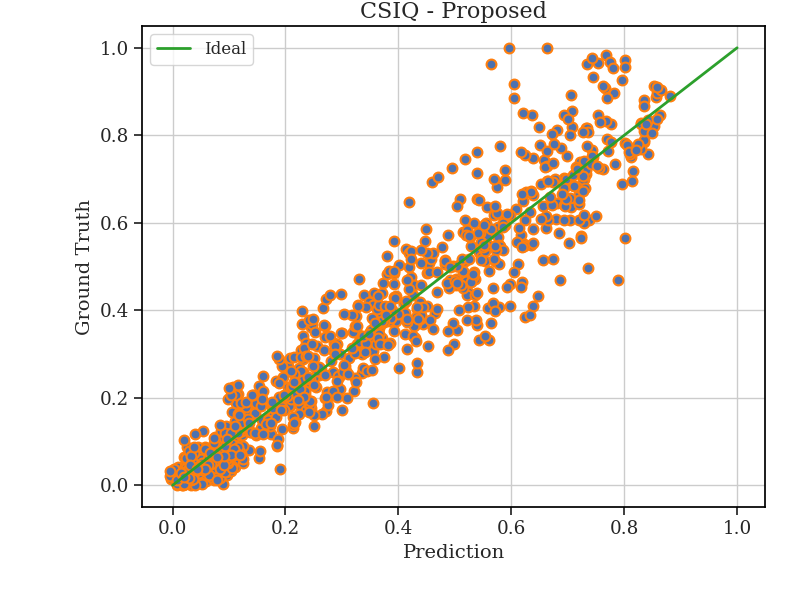}
    \end{subfigure} 

    \begin{subfigure}{0.3\linewidth}
        \includegraphics[width=\linewidth]{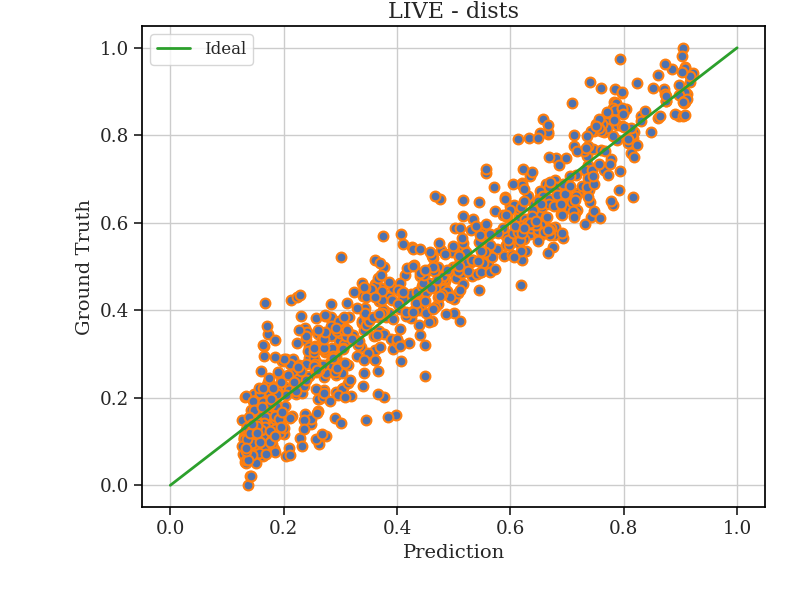}
    \end{subfigure} 
    \begin{subfigure}{0.3\linewidth}
        \includegraphics[width=\linewidth]{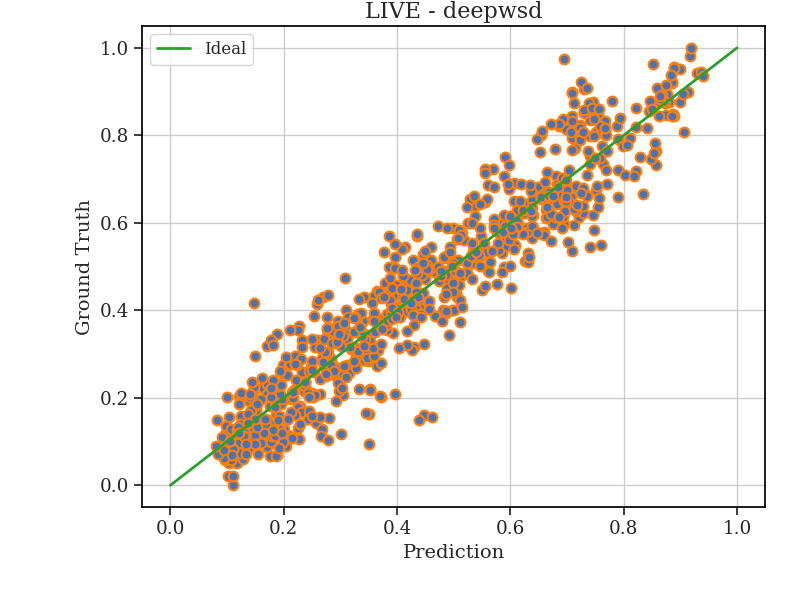}
    \end{subfigure} 
    \begin{subfigure}{0.3\linewidth}
        \includegraphics[width=\linewidth]{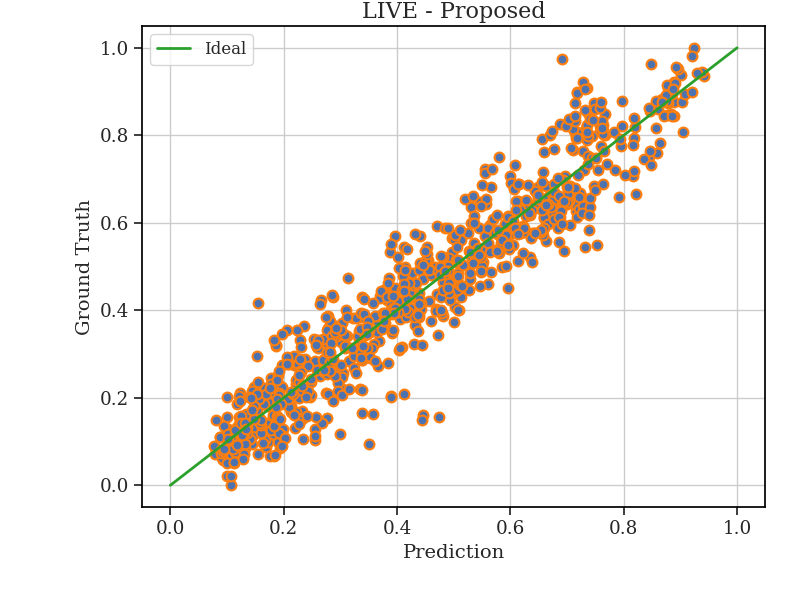}
    \end{subfigure} 
\caption{Visualization of the different methods across databases. }\label{fig03}
\end{figure*}
\begin{table}[!t]
\setlength{\abovecaptionskip}{0pt}%
\setlength{\belowcaptionskip}{9pt}\centering
\caption{\textsc{Evaluated FR-IQA Datasets.}}\label{t1}
\resizebox{0.47\textwidth}{!}{
\centering
\begin{tabular}{cccccc}
\Xhline{1.2pt}\hline\color{black}
Dataset   & LIVE           & CSIQ           & TID2013        & KADID-10k      \\
\hline
\rowcolor{shadegray}Ref No.   & 29             & 30             & 25             & 81             \\
Dist No.  & 779            & 866            & 3000           & 101000         \\
\rowcolor{shadegray}MOS No.   & 25000          & 5000           & 524000         & 30400          \\
Size      & 768$\times$512 & 512$\times$512 & 512$\times$384 & 512$\times$384 \\
\rowcolor{shadegray}Dist Type & Synthetic      & Synthetic      & Synthetic      & Synthetic \\
\hline\hline\Xhline{1.2pt}
\end{tabular}
}
\end{table}
\subsection{Connection with Existing Work and HVS}
Integrating the collaborative feature refinement module and the Hausdorff distance-based distribution similarity measurement module in the proposed framework is inspired by the HVS holistic approach to image quality assessment. HVS considers multiple factors, such as color, luminance, edge, and texture information, when evaluating the perceptual quality of an image. The proposed framework combines the outputs from both modules to provide a comprehensive and accurate image quality assessment that considers the frequency-dependent nature of distortions and the perceptual relevance of the extracted features. The existing FR-IQA methods~\cite{10577432,10.1117/12.597306,7559777} often rely on pairwise comparisons of deep features extracted from reference and distorted images. However, these methods typically neglect the crucial distinction that distortions in color and luminance primarily appear at low frequencies~\cite{NEURIPS2021_f3bd5ad5,Lin_2023_ICCV}. In contrast, distortions in edges and textures are present mainly at high frequencies~\cite{10716315,10533377,10577432}.

The wavelet transform-based collaborative feature refinement module effectively extracts perceptually relevant features. The Hausdorff distance-based distribution similarity measurement module robustly quantifies quality degradation by comparing the feature distributions of the reference and distorted images, ignoring the influence of unexcepted outliers. The HVS inspires the collaborative feature refinement module in the proposed framework to analyse visual information at various scales and orientations. The proposed framework addresses this limitation by employing a wavelet transform to decompose the images into low-frequency and high-frequency subbands, enabling targeted analysis of different distortions. The wavelet transform mimics this behavior by capturing the perceptual information. The refinement function in this module adaptively weights the wavelet subbands on the basis of their perceptual relevance and importance in capturing distortions. This approach is consistent with the HVS sensitivity to different distortions and the ability to prioritize perceptually significant information. The Hausdorff distance-based distribution similarity measurement module is motivated by the HVS tolerance to certain distortion levels while being sensitive to perceptually significant changes. By considering the entire distribution of features rather than relying on point estimates, the Hausdorff distance effectively captures the perceptual quality degradation in a manner that aligns with human perception.

\begin{table*}[!t]
\setlength{\abovecaptionskip}{0pt}%
\setlength{\belowcaptionskip}{9pt}
\caption{FR-IQA prediction comparison in terms of different benchmark datasets. }\label{t2}
\centering
\begin{tabular}{ccccccccccccccccccccccccc}
\Xhline{1.2pt}\hline
\rowcolor{shadegray} Method  & \multicolumn{2}{c}{LIVE} & \multicolumn{2}{c}{CSIQ} & \multicolumn{2}{c}{TID2008} & \multicolumn{2}{c}{TID2013} & \multicolumn{2}{c}{KADID-10k}\\
\cline{2-3}\cline{6-7}\cline{10-11}
        & PLCC  & SRCC  & PLCC  & SRCC  & PLCC    & SRCC  & PLCC    & SRCC  & PLCC  & SRCC  \\\hline

\rowcolor{shadegray} PSNR    & 0.781 & 0.801 & 0.792 & 0.807 & 0.507   & 0.525 & 0.664   & 0.687 & 0.670 & 0.676 \\
SSIM    & 0.847 & 0.851 & 0.810 & 0.833 & 0.625   & 0.624 & 0.665   & 0.627 & 0.610 & 0.619 \\
\rowcolor{shadegray} MS-SSIM & 0.886 & 0.903 & 0.875 & 0.879 & 0.842   & 0.854 & 0.831   & 0.786 & 0.824 & 0.826 \\
VIF     & 0.949 & 0.953 & 0.899 & 0.899 & 0.798   & 0.749 & 0.771   & 0.677 & 0.685 & 0.679 \\
\rowcolor{shadegray} MAD     & 0.904 & 0.907 & 0.922 & 0.922 & 0.681   & 0.708 & 0.737   & 0.743 & 0.717 & 0.726 \\
FSIM    & 0.910 & 0.920 & 0.902 & 0.915 & 0.875   & 0.884 & 0.876   & 0.851 & 0.852 & 0.854 \\
\rowcolor{shadegray} VSI     & 0.877 & 0.899 & 0.912 & 0.929 & 0.871   & 0.895 & 0.898   & 0.895 & 0.877 & 0.878 \\
GMSD    & 0.909 & 0.910 & 0.938 & 0.939 & 0.878   & 0.891 & 0.858   & 0.804 & 0.847 & 0.847 \\
\rowcolor{shadegray} NLPD    & 0.882 & 0.889 & 0.913 & 0.926 & 0.866   & 0.877 & 0.832   & 0.799 & 0.809 & 0.812 \\
LPIPS   & 0.866 & 0.863 & 0.891 & 0.895 & 0.722   & 0.718 & 0.713   & 0.713 & 0.838 & 0.837 \\
\rowcolor{shadegray} DISTS   & 0.924 & 0.925 & 0.919 & 0.920 & 0.829   & 0.814 & 0.854   & 0.830 & 0.886 & 0.886 \\
DeepWSD & 0.904 & 0.925 & 0.941 & 0.950 & 0.900   & 0.904 & 0.894   & 0.877 & 0.887 & \textbf{0.888} \\
\rowcolor{shadegray} Proposed    &  \textbf{0.951}     &  \textbf{0.955}     & \textbf{0.947}      & \textbf{0.963}      &    \textbf{0.892}     &   \textbf{0.911}    &    \textbf{0.895}     &  \textbf{0.891}     &   \textbf{0.888}    & 0.884\\

\hline\hline\Xhline{1.2pt}
\end{tabular}
\end{table*}

\section{Experimental Results}
\subsection{Experimental Settings}
The proposed perceptual-related domain transformation and distribution similarity measurement framework is rigorously validated through comprehensive experiments on several benchmark datasets. Specifically, we assess the proposed measurement via five IQA datasets: TID2008, TID2013, LIVE, CSIQ, and KADID-10k, as detailed in Table \ref{t1}. Our model is benchmarked against various state-of-the-art full-reference IQA methods, including PSNR, SSIM, MS-SSIM, VIF, MAD, FSIM, VSI, GMSD, NLPD, LPIPS, DISTS, and DeepWSD. Each model is evaluated via open-source configurations across the complete datasets. An ablation study examines the impact of patch size on performance. We employ Pearson's linear correlation coefficient (PLCC) and Spearman's rank-order correlation coefficient (SRCC) to evaluate IQA prediction accuracy. A VGG network pretrained on the ImageNet dataset serves as the feature extractor. The implementation is performed via PyTorch, with training and testing on an NVIDIA GTX 4090 GPU. 

\subsection{Performance Evaluation}
The proposed FR-IQA method, which integrates collaborative feature refinement with the Hausdorff distance-based similarity measurement, significantly advances the state-of-the-art methods by addressing specific limitations inherent in existing approaches. As detailed in Table \ref{t2}, our method consistently outperforms traditional metrics such as PSNR and SSIM across multiple benchmark datasets, including LIVE, CSIQ, TID2008, TID2013, and KADID-10k. This superior performance is attributable to the method's ability to address color and luminance distortions in low-frequency domains separately and edge and texture distortions in high-frequency domains, which align closely with the HVS, as highlighted in the abstract. The collaborative feature refinement module employs a wavelet transform that effectively captures multiscale perceptual features, enhancing the extraction of distortions in the wavelet domain, which strongly correlates with subjective human judgments. Figure \ref{fig03} also shows the performance of the proposed method.

Additionally, the Hausdorff distance-based module robustly measures distribution similarities and handles outliers and variations more effectively than conventional distance metrics do, thereby improving the method's resilience across diverse distortion types, as reflected by the consistently high performance on the TID2013 and KADID-10k datasets. Furthermore, the training-free nature of the proposed approach eliminates the dependency on large training datasets, reducing computational overhead while maintaining high accuracy. The extensive evaluation across datasets with varying reference and distortion counts, as outlined in Table \ref{t1}, underscores the method's generalizability and robustness. The proposed method achieves a greater correlation with the HVS. It introduces a novel framework that enhances perceptual quality assessment by leveraging frequency-specific distortions and advanced feature comparison techniques, thereby setting a new benchmark in FR-IQA research. By formulating the FR-IQA problem, we leverage the wavelet transform's multiscale characteristics to capture the image quality information. The statistical analysis of coefficient distributions allows us to quantify the distortions in the image effectively. The proposed approach eliminates the need for a separate training phase using HVS-related parameters on the basis of empirical observations and prior knowledge.

\subsection{Ablation Study}
To assess the impact of different backbone networks on the performance of the proposed FR-IQA method, an ablation study is conducted using VGG, MobileNet, and EfficientNet as backbone architectures, as shown in Table \ref{t3}. The results indicate that the VGG backbone consistently maintains its superiority with the highest PLCC and SRCC values over MobileNet and EfficientNet across all datasets. MobileNet and EfficientNet exhibit lower IQA correlations, indicating reduced effectiveness in capturing the nuanced quality metrics required by these datasets. The inferior performance of MobileNet can be attributed to its lightweight architecture, which may limit its capacity to extract complex perceptual features essential for accurate image quality assessment. Conversely, EfficientNet, though more robust than MobileNet, still falls short of VGG performance because its efficiency over depth and feature extraction richness is optimized. The consistently high performance of VGG across all datasets underscores its effectiveness in feature extraction for FR-IQA tasks. The VGG architecture facilitates the capture of low-frequency distortions, such as color and luminance changes, and high-frequency distortions, including edges and textures, aligning well with the HVS perception mechanisms, as outlined in the abstract. The ablation study demonstrates that the choice of backbone network significantly influences the performance of the proposed FR-IQA method. VGG emerges as a suitable backbone choice among the evaluated options, providing superior correlation metrics across diverse datasets.
\begin{table}[!t]
\setlength{\abovecaptionskip}{0pt}%
\setlength{\belowcaptionskip}{9pt}
\caption{FR-IQA prediction comparison in terms of different backbone networks. }\label{t3}
\centering
\begin{tabular}{ccccccccc}
\Xhline{1.2pt}\hline
\rowcolor{shadegray} Method  &      & VGG  & MobileNet  & EfficientNet  \\\hline
LIVE      & PLCC & \textbf{0.951} & 0.917 & 0.916 \\
          & SRCC & \textbf{0.955} & 0.919 & 0.944 \\\cline{2-5}
CSIQ      & PLCC & \textbf{0.947} & 0.803 & 0.829 \\
          & SRCC & \textbf{0.963} & 0.823 & 0.890 \\\cline{2-5}
TID2008   & PLCC & \textbf{0.892} & 0.674 & 0.714 \\
          & SRCC & \textbf{0.911} & 0.695 & 0.761 \\\cline{2-5}
TID2013   & PLCC & \textbf{0.895} & 0.681 & 0.682 \\
          & SRCC & \textbf{0.891} & 0.712 & 0.771 \\\cline{2-5}
KADID-10k & PLCC & \textbf{0.888} & 0.739 & 0.762 \\
          & SRCC & \textbf{0.884} & 0.760 & 0.821 \\

\hline\hline\Xhline{1.2pt}
\end{tabular}
\end{table}
\begin{table}[!t]
\setlength{\abovecaptionskip}{0pt}%
\setlength{\belowcaptionskip}{9pt}
\caption{FR-IQA prediction comparison in terms of different parameter settings. }\label{t4}
\centering
\begin{tabular}{ccccccccc}
\Xhline{1.2pt}\hline
\rowcolor{shadegray} Method  &      & DWT  & CH  & DWT + CH  \\\hline
LIVE      & PLCC & 0.944 & 0.940 & \textbf{0.951} \\
          & SRCC & 0.952 & 0.950 & \textbf{0.955} \\\cline{2-5}
CSIQ      & PLCC & 0.936 & 0.932 & \textbf{0.947} \\
          & SRCC & 0.961 & 0.960 & \textbf{0.963} \\\cline{2-5}
TID2008   & PLCC & 0.868 & 0.860 & \textbf{0.892} \\
          & SRCC & 0.908 & 0.906 & \textbf{0.911} \\\cline{2-5}
TID2013   & PLCC & 0.866 & 0.857 & \textbf{0.895} \\
          & SRCC & 0.879 & 0.874 & \textbf{0.891} \\\cline{2-5}
KADID-10k & PLCC & 0.850 & 0.843 & \textbf{0.888} \\
          & SRCC & \textbf{0.887} & \textbf{0.887} & 0.884 \\
\hline\hline\Xhline{1.2pt}
\end{tabular}
\end{table}
\subsection{Sensitivity Parameter Analysis}
To evaluate the robustness and effectiveness of different parameter settings in the proposed FR-IQA method, we conducted a sensitivity analysis comparing the discrete wavelet transform (DWT), color histogram (CH), and their combination (DWT + CH) across five benchmark datasets, as presented in Table \ref{t4}. The results consistently demonstrate that the integrated DWT + CH configuration outperforms the individual DWT and CH methods in most cases, aligning with our abstract's assertion that addressing both low-frequency color and luminance distortions alongside high-frequency edge and texture distortions enhances perceptual quality assessment. Specifically, the combination yields the highest PLCC and SRCC on the LIVE and CSIQ datasets, reflecting superior alignment with human visual perception by capturing multiscale perceptual information through wavelet transforms and robust distribution similarity via color histograms. In the TID2008 and TID2013 datasets, the DWT + CH approach significantly improves both the PLCC and the SRCC, underscoring its ability to effectively model diverse distortion types, as highlighted in the proposed perceptual degradation modelling. Although on the KADID-10k dataset, the combined method achieves the highest PLCC but a slightly lower SRCC than individual settings do; however, it still maintains competitive performance, demonstrating the generalizability and resilience of the method across varying image contents and distortion characteristics.
\section{Conclusion}
A novel training-free FR-IQA method that leverages perceptual-related domain transformation and distribution similarity measurement to accurately predict image quality in alignment with HVS is proposed. The wavelet transform-based collaborative feature refinement module effectively extracts perceptually relevant features. The Hausdorff distance-based distribution similarity measurement module robustly quantifies quality degradation by comparing the feature distributions of the reference and distorted images. Integrating these two components enables the proposed method to capture perceptual quality differences without relying on training data or subjective quality scores. The choice of wavelet transform and Hausdorff distance demonstrates strong perceptual relevance and computational efficiency, making the proposed approach adaptable to various IQA scenarios. Extensive experiments conducted on multiple benchmark datasets validate the effectiveness and competitiveness of the proposed method, which achieves state-of-the-art performance and is strongly correlated with human perception.
{
    \small
    \bibliographystyle{ieeenat_fullname}
    \bibliography{main}
}


\end{document}